# Kinetics of structural changes on GaSb(001) singular and vicinal surfaces during the UHV annealing

*A.V. Vasev*[1], M.A. Putyato[1], V.V. Preobrazhenskii[1], A.K. Bakarov[1], A.I. Toropov[1]

[1] Institute of Semiconductor Physics, Siberian Branch of Russian Academy of Sciences, 630090 Novosibirsk, Russia

**Abstract.** The dynamics of processes of antimony desorption was investigated on the singular and vicinal GaSb(001) surface by RHEED method. The role of the terraces edges was determined during antimony evaporation in Langmuir desorption mode. It is shown that the structural transition $(2\times5) \to (1\times3)$ is a complex of two transitions – *order $\to$ disorder* and *disorder $\to$ order*. The influence of the degree of surface miscut from the singular face on the dimension of the transition $(2\times5) \to$ DO was studied. The activation energies of structural transitions $ex(2\times5) \to (2\times5)$, $(2\times5) \to$ DO and DO $\to (1\times3)$ on singular and vicinal faces GaSb(001) were determined.

**Introduction**

The semiconductor compound GaSb (together with InAs and AlSb) belong to the so-called "6.1 Å family", which is mainly interesting to create heterostructures for long wavelength range photonics devices with band gap engineering.

GaSb compound in this family stands alone. It is single one (as opposed to all other compounds III-V) which does not form anion-rich $c(4\times4)$ reconstruction on the fcc(001) surface. Instead, the reconstructions with symmetries $(n\times5)$ and $(n\times3)$ are observed on this surface having cell structures, which are not electro-neutral [1]. Furthermore, on the GaSb(001) surface the bonds of V-V type (Sb-Sb) significantly (~ 1 eV) stronger than bonds of III-V type (Ga-Sb) [1]. The detailed knowledge (at the atomic level) of kinetics of surface processes at typical epitaxy conditions is required when GaSb based structures are creating for photodetector devices applications. The existing data, which are devoted to the study of processes on the GaSb(001) surface, are not enough to create a complete picture.

This work is dedicated to the study of the kinetics of structural changes on the singular and vicinal surfaces GaSb(001) in the process of vacuum annealing by RHEED method.

**1. Experimental**

The researches were carried out on the Compact 21 MBE-machine (*RIBER Co*).

Experiments were realized on the surface of GaSb(001) epitaxial layers with a thickness ~ 500 nm, grown on GaAs(001) wafers, miscut from the singular face by less then 8′ and by 5° in the [110] direction. The singular and vicinal wafers were fixed pairwise (side by side) on the same molybdenum holder using molten indium. Such attachment provides an equivalent thermodynamic conditions during the experiments on these surfaces. The samples orientation is providing the incidence of electrons along the terraces edges (parallel to the direction [−1 10]). The displacement of the electron beam allows to control either the one or the other surface. The procedure of substrate temperature calibration was performed by points of reconstructions transitions $c(4\times4) \to (2\times3)$ and $(2\times3) \to (2\times4)$ on a GaAs(001) [2] and $(2\times5) \to (1\times3)$ on the surface of GaSb(001) [3].

During the experiments the samples were heated under a flux of antimony dimers (BEP($Sb_2$) = $1.55 \cdot 10^{-6}$ Torr) to fixed temperature values in the range of $360 \div 433$°C. The record of specular beam intensity (SBI) of the RHEED patterns is started when the selected temperature was reached and the stationary state was realized. The changes of SBI during desorption were written down after overlap of antimony flux. The GaSb(001) surface was updated by regrowth at 400°C during 5 minutes (V = 0.5 ML/s) before every next measuring.

**2. Results**

The behavior of the singular and vicinal GaSb(001) surfaces under a antimony dimers flux in the low-temperature region ($360 \div 404$°C) is characterized by a clear structural transition $ex(2\times5) \to (2\times5)$ at the same Ts ~ $406 \pm 3$°C. This indicates a negligible impact of terraces edges on the process of antimony dimers leaving from upper (third layer, relative of gallium) layer of $ex(2\times5)$ reconstruction.

The influence becomes significant at high ($409 \div 433$°C) temperatures. After leaving of domains of reconstruction $ex(2\times5)$ from the surface, the dimers vacancy begin to form on the top (now second) antimony layer of the $(2\times5)$ reconstruction on the singular GaSb(001) surface with increasing of temperature. The accumulation of these dimers leads to the formation of areas, which are lost (in terms of RHEED) a long-range order – areas of DO (*disordered*) with unit cell symmetry $(1\times1)$. Dynamics of the DO regions formation is clearly tracked by SBI, and in this case has a distinct character of *order $\to$ disorder* structural transition. On the vicinal surface such changes are not observed. The SBI values for miscut surface remain unchanged in the temperature range $406 \div 433$°C. Thus, increasing of the density of terraces edges due to miscut from the singular face by 5° leads to a shift of the structural transition $(2\times5) \to$ DO into high temperatures area, at least ~ 30°C.

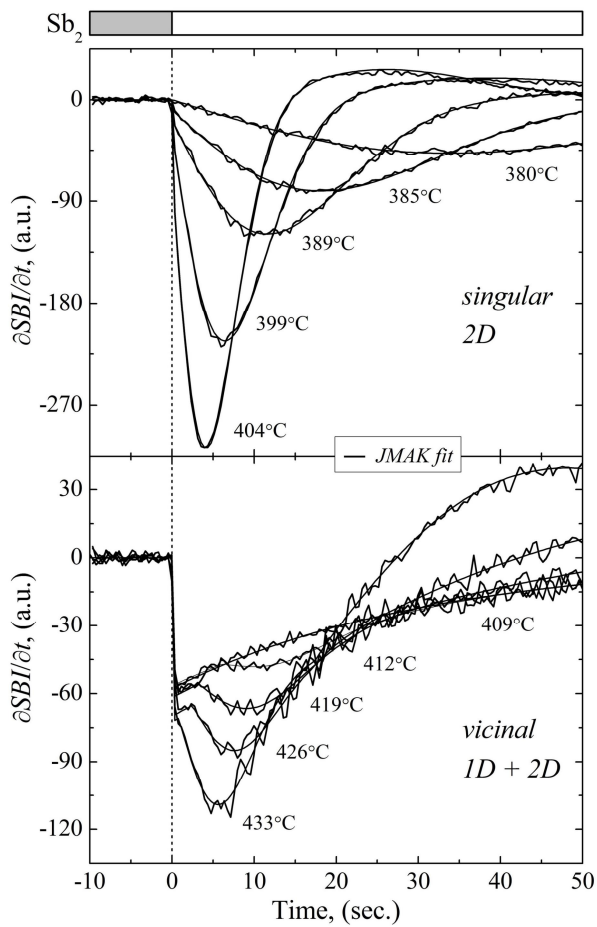

Fig. 1 The analysis of RHEED data within the framework JMAK model.

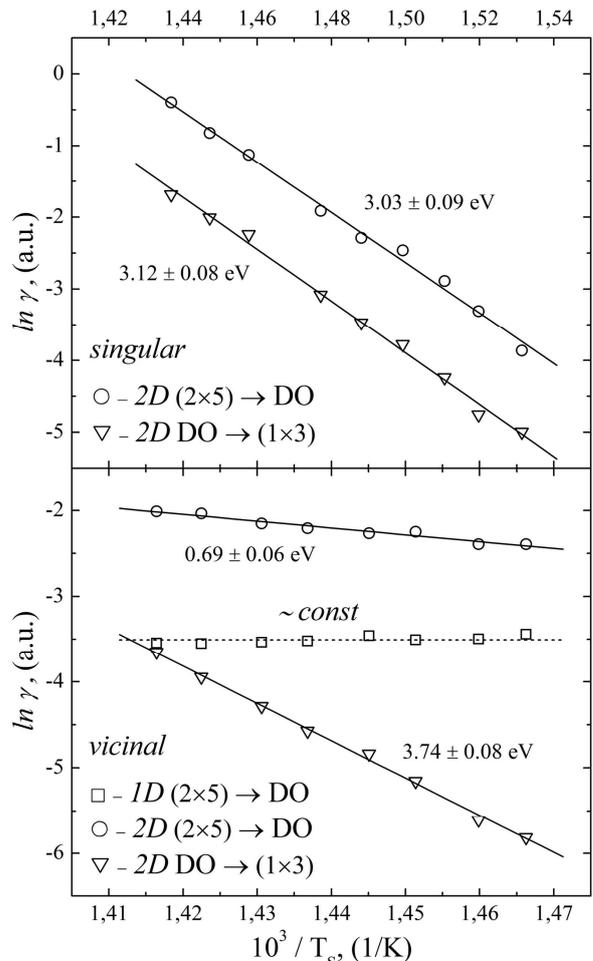

Fig. 2 The activation energies of structural transitions $ex(2\times5) \to (2\times5)$, $(2\times5) \to DO$ and $DO \to (1\times3)$.

The analysis of RHEED data, which were obtained after closing the valve of antimony cell, was realized within the framework JMAK (Johnson-Mehl-Avrami-Kolmogorov) model [4-6] (see Fig. 1). This approach allows to determine the temperature dependence of the velocity ($\gamma$) of structural transitions $ex(2\times5) \to (2\times5)$, $(2\times5) \to DO$ and $DO \to (1\times3)$ on singular and vicinal surfaces, and, as a consequence, to establish their activation energies (see Fig. 2).

It was found that on the vicinal surface the transition $(2\times5) \to DO$ was carried out by two parallel processes – one-dimensional (1D) and two-dimensional (2D), and on the singular surface was realized mainly like two-dimensional process. It is natural to assume, that the reason of this situation is the degree of miscut of the surface from the singular face. To be exact, it is a small width of the terraces on the vicinal surface, which limits the development of a new phase domains in the [110] direction, leaving for growth only direction [$-1$ 10]. An important experimental result is that velocity of the 1D transition $(2\times5) \to DO$ is essentially independent on temperature. Most likely, it is connected with the complex transition scheme. There is a situation where the transition process is limited by the amount of one of the required components, the concentration of which does not change with temperature. Since this is a one-dimensional transition, this component must be somehow associated with the density of the terraces edges. Variants of possible transition mechanisms are discussed.

It was found that the activation energy of the transition $DO \to (1\times3)$ for the singular surface is $3.74 \pm 0.08$ eV, that $\sim 0.6$ eV larger than the value for vicinal surface. This demonstrates the significant influence of the terraces edges on the elastic stresses in the crystal lattice of reconstruction $(1\times3)$. The reconstruction $(1\times3)$ on the singular surface is already stressed, because its structure is not electro-neutral. These stresses can be removed due to the proximity of the terraces edges on vicinal surface.

This work was supported by Russian Science Foundation, grant № 16-12-00023.

**References**
[1] L.J. Whitman, P.M. Thibado, S.C. Erwin, B.R. Bennett and B.V. Shanabrook, Phys. Rev. Lett. **79**(4), 693 (1997).
[2] V.V. Preobrazhenskii, M.A. Putyato, O.P. Pchelyakov and B.R. Semyagin, J. Cryst. Growth. **201/202**, 166 (1999).
[3] M.J. Yang, W.J. Moore, B.R. Bennett and B.V. Shanabrook, Electron. Lett. **34**, 270 (1998)
[4] W.A. Johnson and R.F. Mehl, Trans. Am. Inst. Min. Metall. Eng. **135**, 416 (1939).
[5] M. Avrami, J. Chem. Phys. **9**, 177 (1941).
[6] A.N. Kolmogorov, Bull. Acad. Sci. USSR. Ser. Math. **1**, 355 (1937).